\documentclass[twocolumn]{aastex631}

\usepackage{graphicx}
\usepackage{amsmath}
\usepackage{float}
\usepackage{natbib}
\usepackage{tabularx}
\usepackage{bm}
\usepackage{ulem}
\usepackage{appendix}
\usepackage{longtable}
\usepackage{enumitem}
\usepackage{pifont}

\newcommand\msun{M_{\odot}}

\shorttitle{LMG BHs in Dense Star Clusters}
\shortauthors{Ye et al.}

\graphicspath{{./}{figures/}}

\begin{document}

\title{Lower-mass-gap Black Holes in Dense Star Clusters}

\author[0000-0001-9582-881X]{Claire S. Ye}
\affil{Canadian Institute for Theoretical Astrophysics, University of Toronto, 60 St. George Street, Toronto, Ontario M5S 3H8, Canada}
\correspondingauthor{Claire S.~Ye}
\email{claireshiye@cita.utoronto.ca}

\author[0000-0002-4086-3180]{Kyle Kremer}
\affiliation{Department of Astronomy \& Astrophysics, University of California, San Diego; La Jolla, CA 92093, USA}

\author[0000-0001-5799-9714]{Scott M.\ Ransom}
\affiliation{NRAO, 520 Edgemont Road, Charlottesville, VA 22903, USA}

\author[0000-0002-7132-418X]{Frederic A.\ Rasio}
\affil{Department of Physics \& Astronomy, Northwestern University, Evanston, IL 60208, USA}
\affil{Center for Interdisciplinary Exploration \& Research in Astrophysics (CIERA), Northwestern University, Evanston, IL 60208, USA}

\begin{abstract}

The existence of compact stellar remnants in the mass range $2-5\,\msun$ has long been debated. This so-called lower mass gap was initially suggested by the lack of low-mass X-ray binary observations with accretors about $2-5\,\msun$, but it has recently been called into question following newer observations, including a lower-mass-gap candidate with a millisecond pulsar companion in the dense globular cluster NGC~1851. Here we model NGC~1851 with a grid of similar dense star clusters utilizing the state-of-the-art Monte Carlo $N$-body code \texttt{CMC}, and we specifically study the formation of lower-mass-gap black holes. We demonstrate that both massive star evolution and dynamical interactions can contribute to forming lower-mass-gap black holes. In general, the collapse of massive remnants formed through mergers of neutron stars or massive white dwarfs produces the largest number of lower-mass-gap black holes among all formation channels. However, in more massive clusters, supernova core collapse can contribute comparable numbers. Our NGC~1851-like models can reproduce millisecond pulsar -- lower-mass-gap black hole binaries similar to the observed system. Additionally, the lower-mass-gap black holes can also become components of dynamically assembled binaries, and some will be in merging black hole--neutron star systems similar to the recently detected gravitational wave source GW230529. However, the corresponding merger rate is probably $\lesssim 1~{\rm Gpc^{-3}\,yr^{-1}}$.
\end{abstract}

\section{Introduction} \label{sec:intro}
The maximum mass of neutron stars (NSs) and the transitional mass between NSs and black holes (BHs) have been long-standing questions in astrophysics. Theoretical calculations indicate that NSs have gravitational masses $\lesssim 3\,\msun$ \citep[e.g.,][]{Rhoades_Ruffini_1974,Kalogera_Baym_1996}, while pulsar timing observations have identified $2.08\pm 0.07\,\msun$ \citep{Cromartie+2020,Fonseca+2021} as the most massive NS to date. At the same time, dynamical mass measurements of BHs in X-ray binaries have suggested a lack of BHs with mass $\lesssim 5\,\msun$, indicating the possible existence of a mass gap, usually referred to as the ``lower mass gap'' (henceforth LMG), between $\sim 3$ and $\sim 5\,\msun$ \citep{Bailyn+1998,Ozel+2010,Farr+2011}. These findings inspired searches for the origins of the LMG, including those resulting from core-collapse supernovae \citep{Fryer+2012,Belczynski+2012,Kochanek2014,Liu+2021,Fryer+2022} and binary evolution of massive stars and compact objects \citep[e.g.,][]{Gao+2022,Siegel+2023}.

Recently, however, various observations have revealed several candidates residing in the purported LMG. These include compact objects in non-interacting binaries \citep{Thompson+2019,Jayasinghe+2021}, an isolated compact object detected by microlensing \citep{Lam+2022}, as well as gravitational wave (GW) sources (GW230529 \citealp{GW230529}, GW200115 \citealp{GW200115}, and GW190814 \citealp{gw190814}). These observations suggest that the LMG may not be empty.\footnote{However, previous analyses of the mass spectra of GW detections still found evidence for a paucity of objects in the LMG, without conclusively resolving it \citep[e.g.,][]{GWTC2,Farah+2022}.}

Recently, yet another LMG candidate was identified in the Galactic globular cluster NGC~1851 \citep{Barr+2024}. This object, with inferred mass $2.09-2.71\,\msun$ (95\% confidence interval), was found via radio timing of its companion, the millisecond pulsar PSR J0514‑4002E (NGC~1851E). This could be the most massive NS or least massive BH detected in a millisecond pulsar (MSP) binary to date. MSPs are abundant in globular clusters (GCs) and are understood to form at high rates in these systems due to the role of dynamical processes enabled by high stellar densities. As pointed out by \citet{Barr+2024}, the companion of NGC~1851E bears a striking resemblance in mass to the remnant of the known double neutron star binary merger, GW170817, observed by LIGO-Virgo-KAGRA (LVK; \citealp{GW170817}), potentially hinting that this object formed via a similar mechanism. The double NS merger origin is also motivated by the presence of a known double NS binary with an inspiral time of roughly $200\,$Myr in another Galactic globular cluster, M15 \citep{Jacoby+2006}. Additional formation scenarios are also viable including mergers of massive white dwarf binaries and accretion-induced collapse via binary Roche lobe overflow or tidal disruption of stars \citep[e.g.,][]{Kremer+2021wd,Kremer+2022_nstde,Ye+2024_singlemsp}. Furthermore, scenarios involving single-star evolution cannot be ruled out due to the uncertainties in supernova natal kicks \citep[e.g.,][and references therein]{Odoherty+2023}. However, retention within a GC with central escape velocity $\lesssim100\,{\rm km\,s^{-1}}$ places constraints on the natal kick such an object could have received at birth \citep[e.g.,][and references therein]{Ye+2019}.

It is notable that the host cluster NGC~1851 has one of the highest central densities of Milky Way GCs \citep[e.g.,][]{Baumgardt_Hilker2018}. Indeed, previous studies of the dynamics of NSs in GCs have demonstrated that clusters with high central densities (especially those that have undergone cluster core collapse) feature a markedly enhanced rate in NS dynamics \citep[e.g.,][]{Ye+2019}. In addition to the NGC~1851E binary, clusters that have undergone (or are near to) core collapse are hosts to the aforementioned double NS binary M15C, a population of six apparently young radio pulsars \citep{Boyles_2011,Kremer+2023_youngpulsar}, and an overabundance of single MSPs without binary companions \citep{Ye+2024_singlemsp}, all pointing to processes uniquely enabled by cluster dynamics.

In this study, we explore the formation and dynamics of LMG BHs in dense stellar clusters, similar to the companion of NGC~1851E. We focus in particular upon the formation of these objects via NS collapse, triggered by several mechanisms enabled by dynamical interactions including double NS mergers, white dwarf mergers, and stellar tidal disruptions. We then explore the subsequent dynamical evolution of these mass-gap objects in clusters including their interactions with BHs, NSs, and other stars. We demonstrate that objects like the NGC~1851E binary can form naturally in clusters like NGC~1851 and predict other related GW/electromagnetic sources that may potentially be observed in the future.

We describe the Monte Carlo $N$-body code used for simulating GCs and the compact objects within in Section~\ref{sec:cmc}. We present simulations matching the observed properties of NGC~1851 and describe the grid of core-collapsed cluster simulations in Section~\ref{sec:simulations}. In Section~\ref{sec:formation}, we explore the formation of LMG BHs through massive star evolution and close dynamical encounters in dense star clusters. We study the formation of LMG binaries in Section~\ref{sec:binary}, and potential GW sources and detectable transient signals consisting of LMG BHs in Section~\ref{sec:coll_mer}. Lastly, we discuss the model uncertainties in Section~\ref{sec:discuss} and conclude in Section~\ref{sec:conclu}.

\section{Monte Carlo $N$-body Code and Compact Object Formation}\label{sec:cmc}
We use the Cluster Monte Carlo code (\texttt{CMC}; \citealp[][and references therein]{CMC1}) to study the formation of LMG BHs in GCs. \texttt{CMC} is a Monte Carlo $N$-body code based on the orbit-averaged method of \citet{henon1971monte,henon1971montecluster}. It incorporates various relevant physics for cluster evolution, including two-body relaxation, strong three- and four-body dynamical interactions, and tidal mass loss from the galactic tidal fields. The \texttt{Fewbody} package is used to integrate directly strong three- and four-body gravitational encounters \citep{fregeau2004stellar,fregeau2007monte}, which includes post-Newtonian dynamics for BHs \citep{antognini2014rapid,amaro2016relativistic,Rodriguez+repeated2018,rodriguez2018postb}. Single and binary star evolution is fully coupled to the stellar dynamics and is computed with \texttt{COSMIC} \citep{cosmic}, which is based on \texttt{SSE} \citep{hurley2000comprehensive} and \texttt{BSE} \citep{hurley2002evolution}. 

\subsection{Core-collapse Supernova}\label{subsec:ccsn}
BHs and NSs can form in core-collapse supernovae in \texttt{CMC}. The birth mass distribution of compact objects depends on the supernova mechanism and the formation of LMG BHs may be determined by effects including the explosion energy and the development timescale of instabilities \citep[e.g.,][]{Fryer+2012,Belczynski+2012,Fryer+2022}, or the successfulness of the supernova explosion \citep[e.g.,][]{Kochanek2014}. We follow the prescriptions in \citet{Fryer+2012} for the remnant mass and natal kick distributions of core-collapse supernovae and adopt either their `\textbf{rapid model}' or `\textbf{delayed model}' in the simulations. The delayed mechanism launches an explosion in a longer timescale than the rapid mechanism and allows for a broader range of explosion energy and fallback material onto the proto-NS. This mechanism thus produces a continuous range of remnant masses across the LMG (assuming to be between $2.5\,\msun$ and $5\,\msun$), in contrast to the rapid mechanism which leads to a sharp transition between BHs and NSs and produces very few remnants in the mass gap (Figure~4 in \citet{Fryer+2012} and also the `rapid' and `delayed' histograms in Figure~\ref{fig:bh_ns_mass_tot} below). We adopt a maximum NS mass at $2.5\,\msun$ following, e.g., \citet{Fryer+2012}.\footnote{The maximum mass of NSs is still highly uncertain and depends on their equation of states \citep[e.g.][]{Ozel_Freire_2016}, which allows a maximum mass between about $2\,\msun$ and $3\,\msun$.}

We assume that newly born NSs from core-collapse supernovae receive large natal kicks $v_{\rm NS}$ drawn from a Maxwellian distribution with a velocity dispersion $\sigma_{\rm CCSN} = 265\,{\rm km\,s^{-1}}$ \citep{Hobbs+2005}. The natal kicks are three-dimensional, with their directions sampled isotropically. On the other hand, the natal kicks of BHs are determined by how much material falls back post-explosion and are calculated by 
\begin{equation}
    v_{\rm BH} = (1-f_b)v_{\rm NS},
\end{equation}
where $f_b$ measures the fraction of the stellar envelope mass that falls back upon core collapse. Most BHs in our simulations with massive progenitor stars of $\gtrsim 30\,\msun$ have a considerable amount of fallback materials and are formed with small natal kicks \citep{Fryer+2012}. BHs from low-mass progenitors, however, are assumed to form with a small amount of fallback materials and receive large natal kicks similar to NSs.

\subsection{Electron-capture Supernova and White Dwarf Collapse}\label{subsec:ecsn}
In addition to core-collapse supernovae, NSs can form from electron-capture supernovae in our simulations \citep{nomoto1984evolution,nomoto1987evolution}. Electron-capture can also be triggered in massive white dwarfs (WDs) through accretion onto the WDs or mergers between two WDs \citep{nomoto1991conditions,Saio_Nomoto_1985,Saio_Nomoto_1998,Saio_Nomoto_2004,Shen+2012,Schwab+2015,Schwab+2016,Schwab_2021}. We assume that all super-Chandrasekhar mergers where the components are Carbon-Oxygen (CO) or Oxygen-Neon (ONe) WDs collapse to NSs \citep[e.g.,][]{Kremer+2023_youngpulsar,Ye+2024_singlemsp}, thus maximizing the NS formation rate through this channel. For the natal kicks of electron capture supernovae, we adopt the prescription in \citet{Kiel+2008} and \citet{Kiel_Hurley_2009} and sample the three-dimensional kicks from a Maxwellian distribution with a velocity dispersion $\sigma_{\rm ECSN}=20\,{\rm km\,s^{-1}}$. Therefore, many NSs formed in electron capture supernovae are retained in typical GCs with escape velocities of $\sim 50\,{\rm km\,s^{-1}}$.

Here we follow the treatments of the merger-induced collapses of massive WDs from \citet[][and references therein]{Ye+2024_singlemsp}. We summarize below the two limiting assumptions for bracketing the uncertainties in the mass loss for merger-induced collapses of super-Chandrasekhar WDs: 
\begin{itemize}
    \item \textbf{Default scenario}: As a lower limit, we follow the default mass prescriptions in \citet{cosmic}, where the more massive WD accretes $\Delta M = \dot M_{\rm{edd}}\tau_{\rm{\dot M}}$ from the donor WD during dynamical mass transfer. $\dot M_{\rm{edd}}$ is the Eddington accretion limit, and $\tau_{\rm{\dot M}} = \sqrt{\tau_{\rm{KH}}\tau_{\rm{dyn}}}$ is the characteristic mass-transfer timescale where $\tau_{\rm{KH}}$ is the Kelvin-Helmholtz timescale and $\tau_{\rm{dyn}}$ is the dynamical timescale of the donor WD. This prescription only allows the more massive WD to accrete a small amount of mass from the donor WD, e.g., $\Delta M \approx 10^{-3}\,\msun$ for a $1\,\msun$ WD accreting for $\approx 2000~$yr. The accreting WD in this scenario will collapse to an NS of gravitational mass $0.9\times1.38\,\msun$, where $1.38\,\msun$ is the critical mass for electron capture. 

    \item \textbf{Updated Scenario}: As an upper limit, on the other hand, we assume that the mass loss is negligible during the merging and the subsequent thermal and viscous evolution phase, and the total mass of the pair of WDs is conserved up until collapse \citep{Schwab_2021}. This upper limit allows more WD merger products to exceed the Chandrasekhar mass, thus producing a larger number of NSs and also more massive NSs. Furthermore, if the masses of the merger products exceed the maximum NS mass defined above, we assume that the remnants collapse to a BH.
\end{itemize} 

\subsection{Collapse of Neutron Stars}\label{subsec:nscollapse}
Furthermore, LMG BHs can be produced through dynamical interactions in dense star clusters. NSs engage in frequent dynamical encounters in GCs, which can lead to direct collisions with other stars or compact objects, tidal disruptions of main-sequence stars, or the formation of binaries that could subsequently merge. These collisions and mergers allow the NSs to grow in mass and collapse to BHs if exceeding the maximum NS mass. Specifically, we assume that the remnants of NSs colliding or merging with another compact object conserve the total mass of the merger components in \texttt{CMC}. For collisions with or tidal disruptions of main-sequence stars, we follow the treatments of NS tidal disruption events (TDEs) from \citet{Kremer+2022_nstde} and \citet{Ye+2024_singlemsp} to estimate the accretion rates of disruption debris of the main-sequence stars. 

To briefly summarize, NS-main sequence TDEs occur when the pericenter distance between the two objects during a close encounter is smaller than the tidal disruption radius of the main-sequence star \citep[e.g.,][]{Rees_1988,Perets+2016}:
\begin{equation}
    r_p = \left(\frac{M_{\rm{NS}}}{M_*}\right)^{1/3}R_*\,,
\end{equation}
where $M_{\rm NS}$ is the initial mass of the NS, and $M_*$ and $R_*$ are the initial mass and radius of the main-sequence star, respectively. 

The mass accretion rate onto the NSs during TDEs is calculated as:
\begin{equation}\label{eq:macc}
    \begin{aligned}
        \dot{M}_{acc}&\approx\left(\frac{M_d}{t_v}\right)\left(\frac{R_{acc}}{R_d}\right)^s\\
        &\times\left[1+3(1-C)\left(\frac{t}{t_v}\right)\right]^{-\frac{1+3(1+2s/3)(1-C)}{3(1-C)}}\,,
    \end{aligned}
\end{equation}
where $M_d$, $R_d$, and $t_v$ are the initial mass and radius of the accretion disk, and the viscous accretion time, respectively \citep{Metzger+2008, Kremer+2022_nstde}. The exponent $s \in [0,1]$ is a free parameter determining the amount of mass transported from the outer edge of the accretion disk to the NS \citep[e.g.,][]{Blandford_Begelman_1999}, and $C=2s/(2s+1)$ \citep[e.g.,][]{Kumar+2008}. We assume that $M_d=0.9 M_*$ following the results of the hydrodynamic simulations of \citet{Kremer+2022_nstde}. This is a reasonable assumption since the relative velocity of the NSs and main-sequence stars in typical GCs is, in general, much smaller than the stellar escape velocity. We also assume that $R_d=2r_p$,  $t_v=1~$day, and $R_{acc}$ equals the NS radius \citep{Kremer+2022_nstde}. For more details of the TDE treatments in \texttt{CMC}, see \citet{Ye+2024_singlemsp}.

\section{Modeling Core-collapsed Globular Clusters}\label{sec:simulations}
\subsection{Typical Core-collapse Globular Clusters}\label{subsec:cc-gc}
\begin{figure*}
\begin{center}
\includegraphics[width=\textwidth]{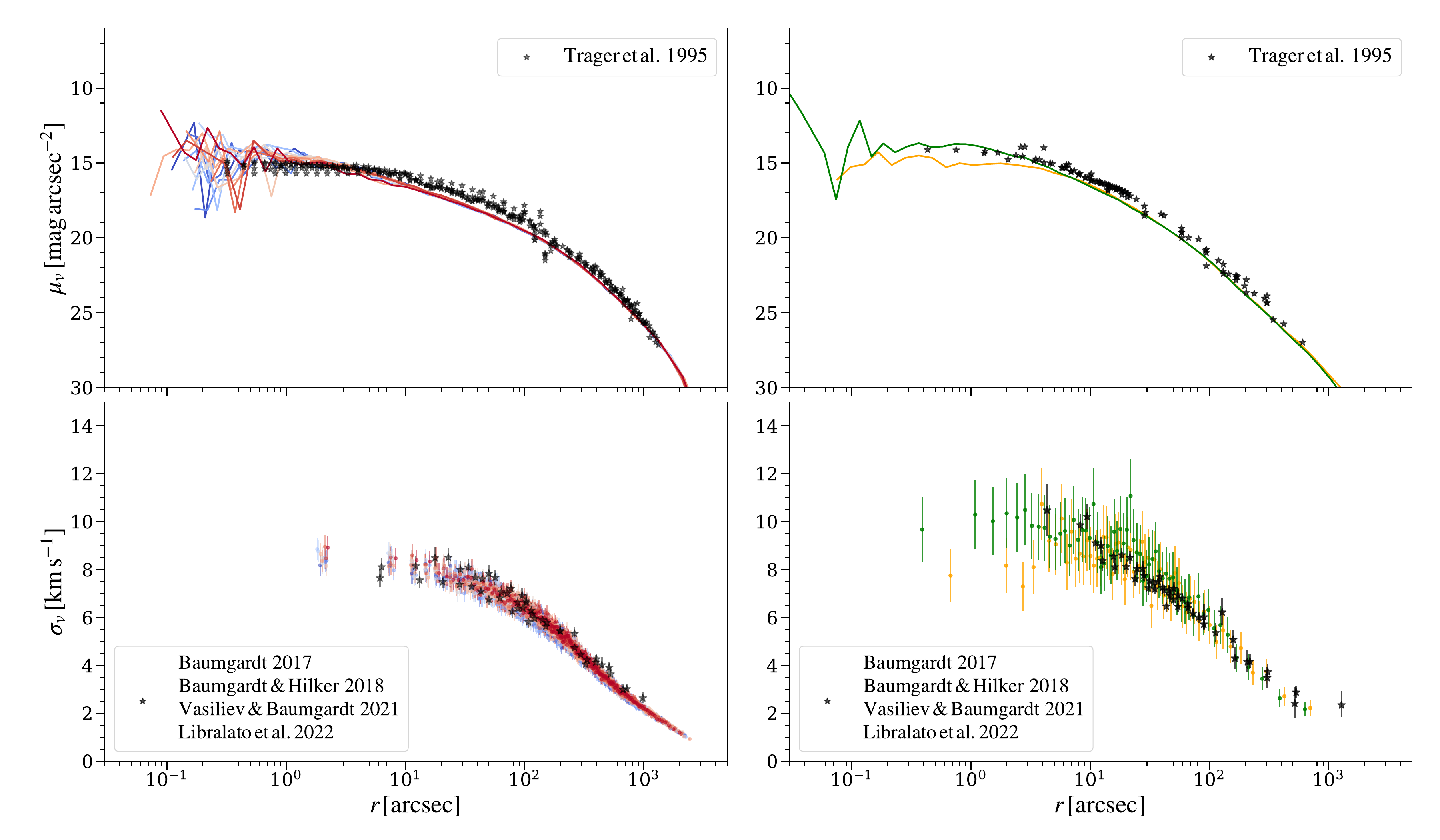}
\caption{Left panels: Comparisons between the observed (black stars) and model (curves and circles with other colors) surface brightness profiles and velocity dispersion profiles of the cluster NGC~6752. The model profiles are for simulation 1-14 in Table~\ref{tab:clu_prop} at $\approx 12~$Gyr. Right panels: Comparisons between the observed (black stars) and model (green and orange curves and circles) surface brightness profiles and velocity dispersion profiles of the cluster NGC~1851. The green profiles (simulation B in Table~\ref{tab:clu_prop}) show the updated model of NGC~1851 at $\approx11~$Gyr (the observed age of the cluster; \citealp{VandenBerg+2013}), while the orange profiles (simulation A) show the default model at $\approx11~$Gyr. The observed surface brightness profile is from \citet{Trager1995}, and the observed velocity dispersion profile is from https://people.smp.uq.edu.au/HolgerBaumgardt/globular/ as well as \citet{Baumgardt_2017}, \citet{Baumgardt_Hilker2018},\citet{Vasiliev_Baumgardt_2021}, and \citet{Libralato+2022}.}\label{fig:sbp_1851}
\end{center}
\end{figure*}

We first run 13 core-collapse GC simulations up to a Hubble time (13.8~Gyr) with the same initial conditions using \texttt{CMC} to explore the formation of LMG BHs through core-collapse supernovae, merger-induced collapse of WDs, and tidal disruption of stars by NSs. These cluster simulations have initial number of stars $N=8\times10^5$, initial virial radius $r_v = 0.5~$pc, metallicity $Z=0.0002$, and Galactocentric distance $r_g=8~$kpc. We adopt the Kroupa standard broken power-law \citep{Kroupa_2001} for the initial mass function between $0.08\,\msun$ and $150\,\msun$. The initial binary fraction is assumed to be $5\%$, representative of the observed binary fraction of old GCs \citep[e.g.,][]{Milone+2012}.\footnote{Studies \citep[e.g.,][]{Fregeau+2009} have shown that the overall hard binary fractions of GCs remain roughly constant with time. In addition, because of the large natal kicks for NSs and LMG BHs, varying the initial binary fraction will likely have negligible effects on the formation of binaries with LMG BHs.} These simulations will evolve to be ``typical'' core-collapsed clusters similar to NGC~6752 in the Milky Way which shows clear evidence for frequent dynamical interactions of lower-mass compact objects \citep[e.g., single MSPs; for more details, see][]{Ye+2024_singlemsp}. Thus clusters similar to NGC~6752 are ideal test grounds for studying the dynamical formation and evolution of LMG BHs. We show the surface brightness profiles and velocity dispersion profiles of all 13 simulations at $\approx 12~$Gyr in Figure~\ref{fig:sbp_1851} (left panels). We focus on core-collapsed GCs in this study since they produce binaries containing NSs, low-mass BHs, or LMG BHs at higher rates than non-core-collapsed GCs \citep[e.g.,][]{Ye+2020_dns}.

For these simulations, we vary the supernova prescriptions, the mass scenarios for WD--WD mergers, and the accretion efficiency of NSs tidally disrupting main-sequence stars (Section~\ref{sec:cmc}). These variations will affect the mass distributions of NSs and the number of LMG BHs formed. We also vary the prescriptions for the tidal capture of two WDs, and the stability criterion of binary mass transfer, $q_{\rm crit}=M_{\rm donor}/M_{\rm accretor}$. This stability criterion will influence the orbital period distribution of the compact object binaries, thereby affecting their cross sections for dynamical encounters \citep[for more details, see][]{Ye+2024_singlemsp}. Table~\ref{tab:clu_prop} lists the different prescriptions adopted for each simulation.

To explore the \textbf{extreme upper limit} for the number of LMG BHs and their binaries formed and retained in GCs, we also run a simulation (simulation~14) which truncates the standard Kroupa initial mass function at $30\,\msun$, resulting in a much smaller number of BHs with masses greater than about $10\,\msun$. Additionally, we assume that NSs born from core-collapse supernovae receive small natal kicks same as those from electron-capture supernovae. These assumptions allow for a large number of NSs and LMG BHs to be retained in the cluster. Other initial conditions of this simulation are the same as mentioned above.

\subsection{Modeling NGC~1851}\label{subsec:ngc1851}
In addition to the main grid of models which are intended to be representative of a ``typical'' core-collapsed Milky Way GC, we also run two additional higher mass simulations designed specifically to match the observed properties of NGC~1851. For these models, we assume an initial number of stars $N=1.3\times10^6$ and virial radius $r_v=0.5$~pc. The metallicity of the simulations is $Z=0.0013$, closely matching the observed metallicity of NGC~1851 from \citet[][2010 edition]{Harris_1996}. We adopt a circular orbit with Galactocentric distance $r_g=20$~kpc for the cluster following https://people.smp.uq.edu.au/HolgerBaumgardt/globular/, which shows that NGC~1851 is on an eccentric orbit in the Milky Way with pericenter distance $r_p\approx0.9$~kpc and apocenter distance $r_a\approx19.9$~kpc. At distances near the pericenter distance, NGC~1851 may be subject to strong tidal stripping from the Milky Way, and a simple circular orbit at the apocenter distance would not be able to capture the detailed evolution of the cluster in the Milky Way's potential. However, simulations with $r_g=20$~kpc match the observed surface brightness profile of NGC~1851 at the outermost part better than those with smaller Galactocentric distances, as we will demonstrate below. This is not surprising since the cluster spends most of its orbital time near the apocenter distance, while only a short time at the pericenter distance. In addition, back scattering--gravitational interactions with other stars scattering the potential escapers back down to a bound orbit--may also prevent a significant fraction of objects from escaping from their host clusters even though they have enough energy to do so \citep[e.g.,][]{Weatherford+2023}. Simulating the detailed evolution of the cluster's tidal boundary is beyond the scope of this paper, and the location of the tidal boundary likely does not have a significant effect on the LMG BHs, which mostly concentrate at the cluster cores due to mass segregation. Other initial cluster properties (e.g., binary fraction, initial mass function, stellar distribution profile) are the same as the simulations in the main grid (simulations 1-13 above). For these two simulations, we again adopt different prescriptions that will affect the mass of the NSs and BHs for comparison (simulations~A and B in Table~\ref{tab:clu_prop}). Furthermore, given the observed eccentric orbit of NGC~1851, we include simulation~C with Galactocentric distance $r_g=1$~kpc as a limiting case. All the other initial conditions of this simulation are the same as in simulation~B.

\begin{deluxetable*}{c|cccccc|ccccc}
\tabletypesize{\scriptsize}
\tablewidth{-1pt}
\setlength{\tabcolsep}{1.5pt}
\tablecaption{Simulation Results for Core-collapsed Clusters} \label{tab:clu_prop}
\tablehead{
\multicolumn{1}{c|}{Model} & \colhead{$M_{clu}$} & \colhead{$N_{\rm{BH}}$} & \colhead{$N_{\rm{NS}}$} & \colhead{$N_{\rm{NS-WD}}$} & \colhead{$N_{\rm{NS-NS}}$} & \multicolumn{1}{c|}{$N_{\rm{BH-NS}}$}& \colhead{SN} & \colhead{$M^{\rm WD}_{\rm scenario}$} & \colhead{TDE} & \colhead{WDTC} & \colhead{qcrit}\\
\multicolumn{1}{c|}{} & \colhead{[$10^5\msun$]} & \colhead{} & \colhead{} & \colhead{} & \colhead{} & \multicolumn{1}{c|}{} & \colhead{} & \colhead{} & \colhead{} & \colhead{} & \colhead{}
}
\startdata
1 & 1.97 & 5 & 968 & 18 & 1 & 0 & rapid & default & - & - & default\\
$2^d$ & 1.90 & 2 & 905 & 17 & 2 & 0 & delayed & default & - & - & default\\
$3^m$ & 1.99 & 7 & 917 & 21 & 3 & 1 & rapid & updated & - & - & default\\
$4^{dm}$ & 1.89 & 5 & 862 & 16 & 0 & 0 & delayed & updated & - & - & default\\
\hline
5 & 2.01 & 3 & 889 & 17 & 2 & 0 & rapid & default & 0.2 & \checkmark & default\\
$6^d$ & 2.02 & 3 & 970 & 20 & 1 & 0 & delayed & default & 0.2 & \checkmark & default\\
$7^m$ & 1.97 & 12 & 904 & 12 & 0 & 0 & rapid & updated & 0.2 & \checkmark & default\\
$8^{dm}$ & 2.04 & 9 & 937 & 21 & 2 & 1 & delayed & updated & 0.2 & \checkmark & default\\
\hline
9 & 2.07 & 4 & 961 & 22 & 1 & 0 & rapid & default & 0.2 & \checkmark & updated\\
$10^d$ & 1.96 & 1 & 1077 & 13 & 1 & 0 & delayed & default & 0.2 & \checkmark & updated\\
$11^m$ & 2.01 & 15 & 955 & 21 & 2 & 1 & rapid & updated & 0.2 & \checkmark & updated\\
$12^{dm}$ & 1.98 & 10 & 968 & 12 & 2 & 0 & delayed & updated & 0.2 & \checkmark & updated\\
\hline
$13^{dm}$ & 2.05 & 10 & 905 & 16 & 1 & 1 & delayed & updated & 0.0 & \checkmark & updated\\
$14^{dm*}$ & 2.03 & 13 & 3181 & 10 & 6 & 1 & delayed & updated & 0.2 & \checkmark & updated\\
\hline\hline
A & 3.79 & 17 & 816 & 20 & 0 & 0 & rapid & default & - & - & default\\
$B^{dm}$ & 3.80 & 24 & 970 & 22 & 1 & 1 & delayed & updated & 0.2 & \checkmark & updated\\
\hline
$C^{dm*}$ & 2.23 & 15 & 936 & 20 & 2 & 1 & delayed & updated & 0.2 & \checkmark & updated

\enddata
\tablecomments{From left to right, the columns are the total mass of the cluster at present, the number of BHs retained at present, the number of NSs retained at present, and the number of compact object binaries (NS--WD, NS--NS, and BH--NS) at present. The present-day values for simulations 1-14 are averages between 11 and 13.8~Gyr in each cluster simulation, and between 10 and 13~Gyr for simulations~A, B, and C.\\ 
--The last five columns indicate the various physical processes included in the simulations and relevant to forming LMG BHs. `SN' shows the supernova model used. `$M^{\rm WD}_{\rm scenario}$' shows the mass scenario assumed for the merger-induced collapse of two WDs (Section~\ref{sec:cmc}). `TDE' indicates whether the simulations include the tidal disruption of main-sequence stars by NSs and the subsequent accretion of the disrupted debris by the NSs (see also Section~\ref{subsec:nscollapse}). The numbers show the $s$-parameter values assumed for Equation~\ref{eq:macc}. $s=0$ corresponds to the highest accretion efficiency. `WDTC' indicates if WD--WD tidal capture is turned on in the simulations. The inclusion of WD--WD tidal capture can slightly increase the WD merger rates. `qcrit' shows the critical mass prescription used for stable mass transfer (default: \citealp{hurley2000comprehensive} and \citealp{Hjellming_Webbink_1987}; updated: \citealp{Belczynski+2008}). The updated prescription for `qcrit' allows for more stable mass transfer, thus generally producing wider binaries that are easier to disrupt by gravitational encounters. For more details, see \citet{Ye+2024_singlemsp}.\\
--Note that simulation~14 is an \textbf{extremely optimistic} model where most NSs and LMG BHs receive small natal kicks and are retained in the cluster. The initial mass function of this model is also truncated at $30\,\msun$ to prevent the formation of massive BHs.\\
--Simulation~C has the same initial conditions as simulation~B except for the Galactocentric distance $r_g$. Simulation~B has $r_g=20$~kpc while simulation~C has $r_g=1$~kpc.}
\end{deluxetable*}

Figure~\ref{fig:sbp_1851} (right panels) shows comparisons between the model and observed surface brightness profiles and velocity dispersion profiles of NGC~1851. We exclude very luminous stars with luminosity $L>12\,L_{\odot}$ to avoid large fluctuations in the model surface brightness profiles, and we calculate the model velocity dispersion profiles with only the giant stars and upper main-sequence stars with $M>0.8\,\msun$ (the main-sequence turnoff mass of the simulations is $\sim0.85\,\msun$). The orange profiles correspond to simulation A in Table~\ref{tab:clu_prop} (which assumes the rapid supernova prescription and the default mass scenario for WD--WD mergers). The green profiles are for simulation B in the same table (which assumes the delayed supernova prescription and the updated WD mass scenario). Both simulations produce similar surface brightness and velocity dispersion profiles at the present day because of the same initial conditions used. We note that the model surface brightness profiles do not overlap exactly with the observed ones. However, our goal is not to produce a best-fit NGC~1851 model in this study, but to explore the formation pathways of binaries similar to the NGC~1851E binary in models that match the observed properties of NGC~1851 reasonably closely.

Table~\ref{tab:clu_prop} summarizes the present-day properties of the 14+2 GC simulations and their prescriptions for forming LMG BHs. The present-day values are averaged between 11 and 13.8~Gyr for simulations 1-14, while between 10 and 13~Gyr for simulations A-B (the observed age of NGC~1851 is $\sim 11$~Gyr; \citealp{VandenBerg+2013}). All clusters have reached core-collapsed at late times, which is shown by the small number of BHs retained \citep[e.g.,][and references therein]{Kremer+2020bhburning}. In the following sections, we will explore the contributions from various pathways to the formation of binaries and transient sources containing LMG BHs.

\section{Formation of Lower-mass-gap Objects}\label{sec:formation}
In this section, we quantify the effects of the rapid and the delayed models of core-collapse supernovae from \citet{Fryer+2012} (Section~\ref{subsec:ccsn}), accretion- and merger-induced collapses of WDs (Section~\ref{subsec:ecsn}), as well as the collapses of NSs through mergers and TDEs (Section~\ref{subsec:nscollapse}), on the population of LMG BHs in GCs.

\begin{figure*}
\begin{center}
\includegraphics[width=\textwidth]{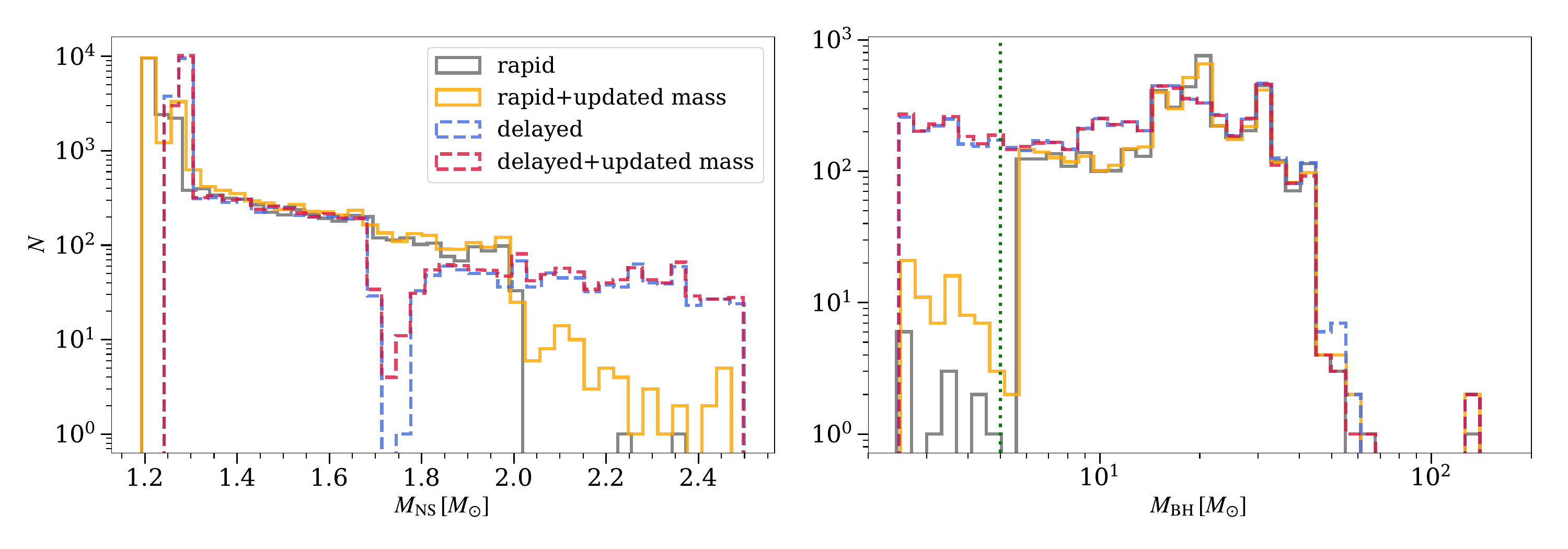}
\caption{Mass distributions of NSs and BHs formed throughout the evolution of the clusters (simulations 1-12 in Table~\ref{tab:clu_prop}) over a Hubble time, including BHs and NSs resulting from the collapse of massive NSs and WDs from collisions/mergers, respectively. The vertical line shows $5\,\msun$, the upper boundary of the LMG assumed here. `Rapid' and `delayed' refer to the rapid and delayed supernova models in \citet{Fryer+2012}, respectively. `Updated mass' refers to the updated scenario which conserves mass during WD--WD merger and the subsequent collapse of the remnant (as opposed to the default scenario which does not conserve mass; Section~\ref{sec:cmc}). The delayed supernova prescription produces more objects between $2\,\msun$ and $5\,\msun$ compared to the rapid prescription.}\label{fig:bh_ns_mass_tot}
\end{center}
\end{figure*}

Figure~\ref{fig:bh_ns_mass_tot} shows the mass distributions of BHs and NSs formed over a Hubble time in simulations 1-12. The simulations are divided into four groups that utilize the rapid and delayed supernova prescriptions \citep{Fryer+2012} as well as the default and updated mass scenarios for the mergers between a pair of WDs (Table~\ref{tab:clu_prop}). Simulations that adopt the delayed supernova prescription produce a significant number of objects in the LMG ($\sim 465$ per cluster over a Hubble time), in contrast to the ones using the rapid prescription ($\sim 4$ per cluster over a Hubble time). 

\begin{figure}
\begin{center}
\includegraphics[width=\columnwidth]{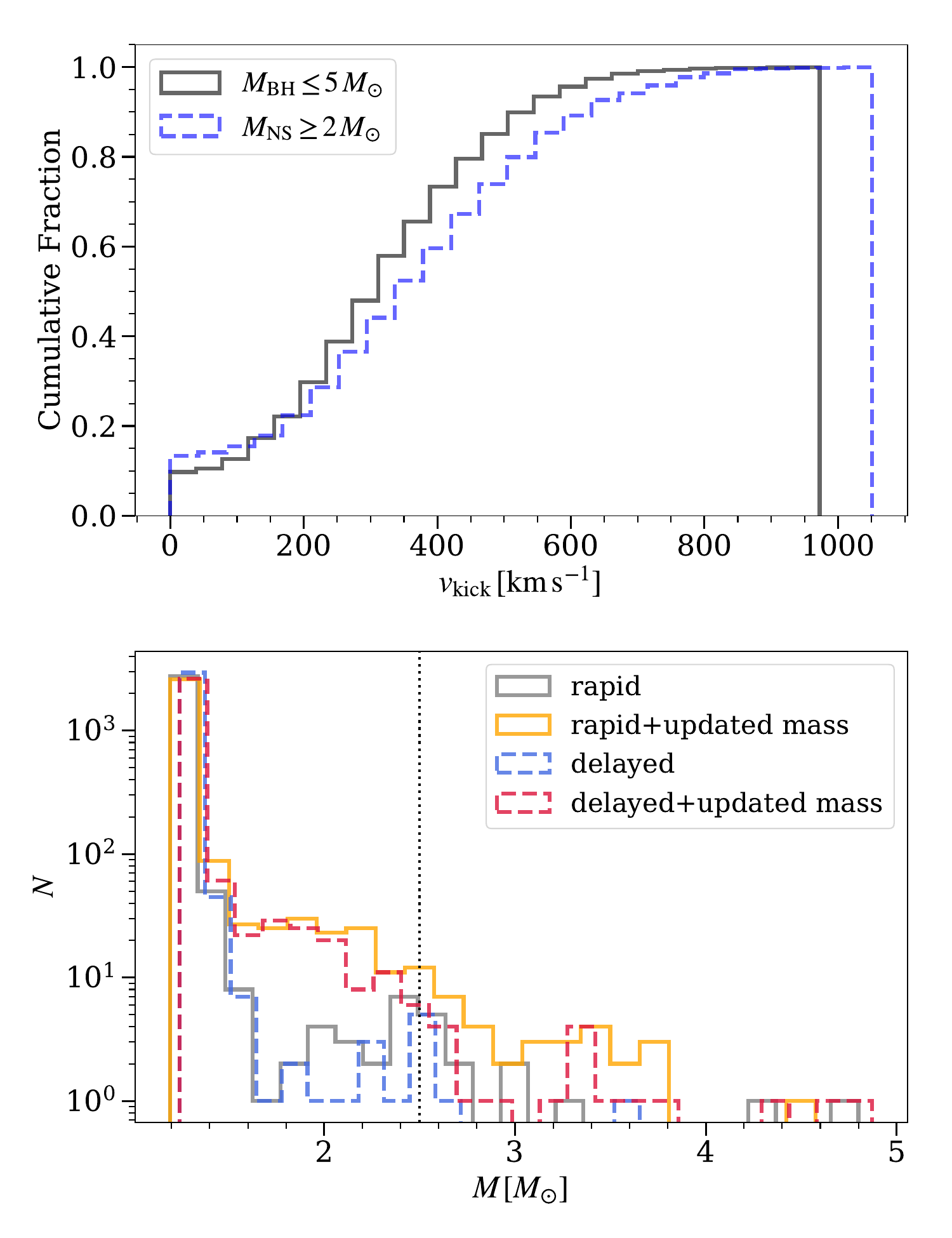}
\caption{Upper panel: natal kick distributions of newly born NSs with masses $\geq2\,\msun$ and BHs with masses $\leq5\,\msun$ for simulations 1-12 from Table~\ref{tab:clu_prop}. Only $\sim 10\%$ of these NSs and BHs have small enough natal kicks to be retained in the clusters. Bottom panel: mass distributions of NSs and low-mass BHs at present ($\approx 12$~Gyr) for the same 12 simulations. The vertical line marks $2.5\,\msun$, the maximum NS mass adopted in this study. The simulations with the rapid and the delayed supernova prescriptions have similar present-day NS and BH mass distributions, while the simulations adopting the updated mass scenario for WD merger have more massive NSs and low-mass BHs.}\label{fig:bh_ns_mass_present}
\end{center}
\end{figure}

Despite the large number of LMG BHs born, most of them are not retained in the host clusters. In our simulations (except the extremely optimistic simulation~14), NSs up to $2.5\,\msun$ receive large natal kicks of hundreds of ${\rm km\,s^{-1}}$, and BHs smaller than about $5\,\msun$ receive fallback kicks at birth similar to the NS kicks (Section~\ref{sec:cmc}). The upper panel of Figure~\ref{fig:bh_ns_mass_present} shows the natal kick distributions for massive NSs and low-mass BHs in simulations 1-12. More than $\sim 90\%$  of the compact objects between $2$ and $5\,\msun$ have natal kicks $\gtrsim 50\,{\rm km\,s^{-1}}$. Since the escape velocities of typical GCs are generally $\lesssim50-100\,{\rm km\,s^{-1}}$, most of the LMG BHs formed from the delayed prescriptions are ejected from their host clusters at birth. This leads to the retention of only a few LMG BHs from the delayed supernovae. The bottom panel of Figure~\ref{fig:bh_ns_mass_present} shows the mass distributions of NSs and BHs of the same 12 simulations at the present day. The similarity of the remnant mass distributions between the simulations using the rapid and delayed prescriptions demonstrates the negligible effects of the supernova prescriptions on the LMG populations in typical GCs, \textbf{if LMG remnants have large natal kicks}. Table~\ref{tab:clu_coll_mer} also shows the number of LMG BHs formed in supernovae that are retained in the clusters ($\sim 5$ per cluster for typical Milky Way core-collapsed clusters; simulation 1-13). If all NSs and LMG BHs receive small natal kicks, however, hundreds of LMG BHs can be retained in a cluster (simulation~14). For more massive clusters similar to NGC~1851 (simulations A and B), the number of retained LMG BHs is larger due to the cluster's higher escape velocity.

Columns 2-5 in Table~\ref{tab:clu_coll_mer} show the contributions to the formation of LMG BHs from various channels. These numbers only include the retained remnants and the mergers in the clusters. Overall, mergers between NSs and WDs produce the largest number of LMG BHs compared to other channels (except when TDE accretion is almost conservative and when all NSs and LMG BHs receive low natal kicks), with the numbers comparable to or much larger than the combined total from other channels. Those involving two NSs contribute to $\lesssim 50\%$ of the numbers formed by the NS--WD merger channel in most cases, with variations for different simulations. This is not surprising since the number of massive WDs in GCs is significantly larger than that of NSs from stellar evolution, even though NSs are more massive. 

For simulations 1-13 that adopt the delayed supernova prescription, $\sim 5$ LMG BHs are retained. This number can be comparable to that from the NS--NS merger channel in simulations with the default mass scenario for WD mergers, but it is much smaller than either the NS--WD or NS--NS channel when the updated mass scenario is adopted. On the other hand, for more massive clusters (e.g., clusters like NGC~1851; simulation B) the number of retained LMG BHs formed through delayed supernovae can be several times higher due to the larger cluster escape velocity, resulting in a contribution similar to that of the NS--WD merger channel. We can expect that in even more massive clusters or with smaller natal kicks, more LMG BHs could be produced by supernovae rather than forming through the collapse of NSs from dynamical interactions. We will explore different supernova prescriptions and compact object natal kicks in future studies.

NS-main-sequence star TDEs generally have minor contributions to forming LMG BHs even with relatively high accretion efficiency of the tidally disrupted debris (Table~\ref{tab:clu_coll_mer}). Simulations 5-12 assume that $s=0.2$ in Equation~\ref{eq:macc}, corresponding to $\sim 0.1\,\msun$ accreted onto the NS in $\sim 10^8$~seconds from the disruption of a $1\,\msun$ main-sequence star \citep{Kremer+2022_nstde}. This amount of mass is insufficient to cause NSs with typical masses of about $1.3\,\msun$ to collapse to BHs. Only with the largest accretion rate when $s=0$ (corresponding to $\sim 1\,\msun$ accreted in $\sim 10^8$~seconds from the disruption of a $1\,\msun$ main-sequence star) can many NSs collapse to low-mass BHs from the tidal disruption channel. This is demonstrated in Table~\ref{tab:clu_coll_mer} by simulation 13, where the TDE channel becomes the dominant channel for producing LMG BHs.

\begin{deluxetable*}{c|cccc|cccccccc|cccc}
\tabletypesize{\scriptsize}
\tablewidth{-1pt}
\setlength{\tabcolsep}{1.5pt}
\tablecaption{Collisions and Mergers of Compact Objects in Core-collapsed Clusters} \label{tab:clu_coll_mer}
\tablehead{
\multicolumn{1}{c|}{} & \multicolumn{4}{c|}{LMG BHs from} & \colhead{} & \colhead{} & \colhead{} & \colhead{} & \colhead{} & \colhead{} & \multicolumn{2}{c|}{Frac. with LMG BHs} & \colhead{} & \colhead{} & \colhead{} & \colhead{Frac.}\\
\multicolumn{1}{c|}{Model} & \colhead{$N_{\rm delayed}$} & \colhead{$N_{\rm TDE}$} & \colhead{$N_{\rm NSNS}$} & \multicolumn{1}{c|}{$N_{\rm NSWD}$} & \colhead{$N_{\rm{BHBH}}$} & \colhead{$N_{\rm{NSNS}}$} & \colhead{$N_{\rm{BHNS}}$} & \colhead{$N_{\rm{BHWD}}$} & \colhead{$N_{\rm{NSWD}}$} & \colhead{$N_{\rm{WDWD}}$} & \colhead{$F_{\rm mg}^{\rm BHNS}$} & \multicolumn{1}{c|}{$F_{\rm mg}^{\rm BBH}$} & \colhead{$N_{\rm WDMS}$} & \colhead{$N_{\rm NSMS}$} & \colhead{$N_{\rm BHMS}$} & \colhead{$F_{\rm mg}^{\rm BHMS}$}
}
\startdata
1 & 0 & 0 & 4 & 6 & 149 & 6 & 3 & 12 & 102 & 648 & 1.00 & 0.007 & 1289 & 134 & 120 & 0.05\\
$2^d$ & 7 & 0 & 6 & 15 & 146 & 11 & 7 & 24 & 126 & 675 & 0.29 & 0.034 & 1999 & 226 & 165 & 0.10\\
$3^m$ & 0 & 0 & 6 & 38 & 153 & 14 & 13 & 28 & 103 & 603 & 0.69 & 0.026 & 1410 & 176 & 141 & 0.10\\
$4^{dm}$ & 2 & 0 & 17 & 44 & 144 & 28 & 18 & 32 & 142 & 586 & 0.83 & 0.062 & 1873 & 248 & 227 & 0.16\\
\hline
5 & 0 & 0 & 4 & 9 & 146 & 8 & 5 & 14 & 95 & 652 & 0.40 & 0.007 & 1378 & 150 & 232 & 0.03\\
$6^d$ & 6 & 5 & 5 & 11 & 131 & 14 & 7 & 31 & 121 & 740 & 0.57 & 0.023 & 1638 & 224 & 281 & 0.05\\
$7^m$ & 0 & 2 & 14 & 45 & 145 & 17 & 8 & 19 & 112 & 615 & 0.75 & 0.028 & 1424 & 214 & 308 & 0.14\\
$8^{dm}$ & 5 & 1 & 20 & 40 & 126 & 26 & 17 & 35 & 135 & 643 & 0.82 & 0.024 & 1508 & 222 & 252 & 0.13\\
\hline
9 & 0 & 1 & 7 & 8 & 133 & 12 & 6 & 8 & 80 & 336 & 0.67 & 0.008 & 1331 & 143 & 268 & 0.03\\
$10^d$ & 2 & 3 & 11 & 18 & 123 & 17 & 9 & 28 & 132 & 411 & 0.78 & 0.016 & 2084 & 334 & 321 & 0.10\\
$11^m$ & 0 & 1 & 16 & 37 & 143 & 16 & 12 & 25 & 92 & 358 & 0.83 & 0.028 & 1322 & 176 & 301 & 0.11\\
$12^{dm}$ & 2 & 2 & 20 & 43 & 129 & 26 & 18 & 37 & 105 & 372 & 0.72 & 0.078 & 1769 & 307 & 328 & 0.11\\
\hline
$13^{dm}$ & 6 & 56 & 10 & 34 & 142 & 19 & 22 & 33 & 74 & 354 & 0.86 & 0.127 & 1582 & 249 & 302 & 0.20\\
$14^{dm*}$ & 315 & 18 & 50 & 56 & 220 & 82 & 86 & 34 & 118 & 160 & 0.78 & 0.377 & 1313 & 821 & 678 & 0.46\\
\hline\hline
A & 0 & 0 & 0 & 0 & 256 & 0 & 1 & 7 & 16 & 449 & 0.00 & 0.000 & 649 & 25 & 183 & 0.00\\
$B^{dm}$ & 19 & 1 & 5 & 21 & 254 & 2 & 14 & 21 & 45 & 308 & 0.57 & 0.031 & 1740 & 210 & 437 & 0.08\\
\hline
$C^{dm*}$ & 24 & 5 & 20 & 55 & 246 & 21 & 22 & 37 & 91 & 396 & 0.95 & 0.057 & 2440 & 318 & 535 & 0.20

\enddata
\tablecomments{Columns 2-5 show the delayed supernova formation and the three dynamical formation pathways of BHs in the LMG ($2.5-5\,\msun$). From left to right, it is the number of LMG BHs from the delayed supernova prescription, number of NS collapsing to BHs through NS-main-sequence star TDEs, binary NS mergers, and NS--WD mergers, respectively.\\
--Columns 6-11: the number of various mergers (binary BHs, binary NSs, BH--NS, BH--WD, NS--WD, and binary WDs) in each simulation over a Hubble time. Columns 12-13 show the fraction of BH--NS and binary BH mergers that contain at least one LMG BHs.\\
--Columns 14-16 are the collisions and tidal disruptions between WDs and main-sequence stars, NSs and main-sequence stars, as well as BHs and main-sequence stars, respectively. The last column shows the fraction of BH-main-sequence star collisions/TDEs that involve a BH in the LMG.\\
--Simulation~14 shows an \textbf{extreme} case where all NSs and LMG BHs receive small natal kicks the same as those from electron-capture supernovae, and the initial mass function of the simulation is truncated at $30\,\msun$.\\
--Simulation~C shows a limiting case with Galactocentric distance $r_g=1$~kpc. Other initial conditions are the same as in simulation~B.}
\end{deluxetable*}

\section{Lower-mass-gap Binaries}\label{sec:binary}
Having established that LMG BHs can form in GCs through various mechanisms, we now explore the subsequent dynamical evolution of these objects, in particular the dynamical formation of LMG binaries. In general, only a very small number of NS--NS and BH--NS binaries are present at late times for all cluster simulations, as shown in Table~\ref{tab:clu_prop} (average of 6 and up to about 10 in the extremely optimistic simulation~14). This is roughly consistent with the fact that, despite the hundreds of MSPs observed in Milky Way GCs \citep[and found in simulations, e.g.,][]{Ye+2020_dns,Ye+2022_47tuc}, only two pulsar-NS (or low-mass BH) binaries have been definitively identified. Both are in massive, core-collapsed, or near core-collapsed clusters (M15 and NGC 1851). We list the properties of the MSP NGC~1851E and its binary in Table~\ref{tab:ns_binary}, together with those of the confirmed or candidate NS--NS binaries in all Milky Way GCs for reference.

\begin{figure*}
\begin{center}
\includegraphics[width=\textwidth]{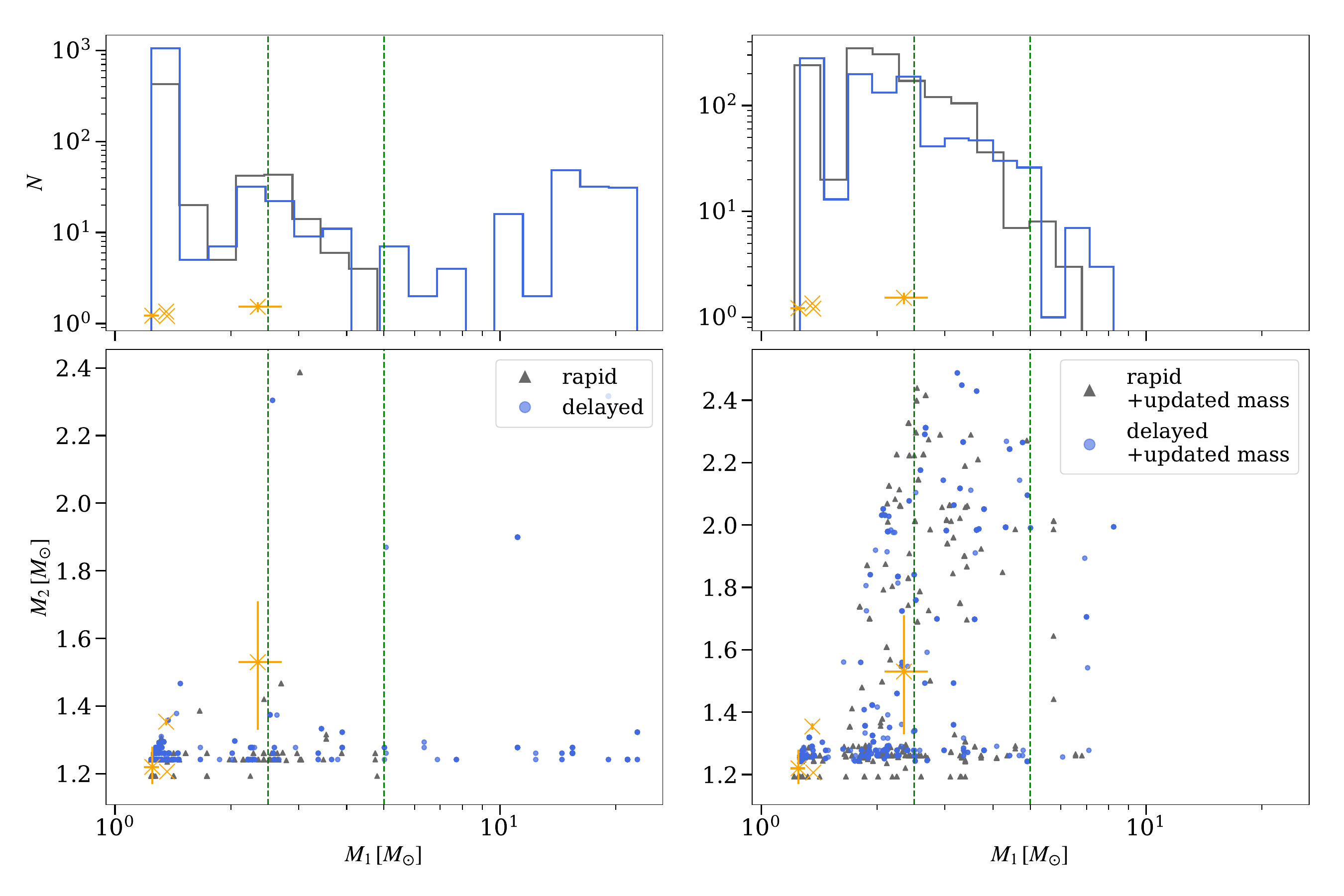}
\caption{Component masses of all NS--NS and BH--NS binaries (bottom panels) and the mass distributions of their primaries (upper panels) between about 11 and 14~Gyr for the core-collapsed cluster simulations 1-12 in Table~\ref{tab:clu_prop}. Blue histograms and markers are for simulations with the rapid supernova prescription and black are for those with the delayed supernova prescription. The panels on the left show simulations that adopt the default mass scenario for WD--WD mergers while the ones on the right adopt the updated mass scenario (the latter assumes mass conservation during the mergers; section~\ref{subsec:ecsn}). The orange crosses show the pulsar binaries and candidates listed in Table~\ref{tab:ns_binary} (except for the M30B binary which only has upper and lower limits for the component masses).}\label{fig:binary_mass}
\end{center}
\end{figure*}

However, different mass prescriptions for compact objects can significantly alter the present-day mass distribution of NS--NS and BH--NS binaries, as demonstrated in Figure~\ref{fig:binary_mass} for simulations 1-12. The delayed supernova prescription is more likely to produce BH--NS binaries with BH mass $>5\,\msun$ compared to the rapid prescription. Still, both supernova prescriptions yield similar numbers of binaries with massive NSs and LMG BHs (Figure~\ref{fig:binary_mass}, upper left panel). The most significant change in the BH/NS--NS binary mass distribution arises from the updated mass scenario for WD merger and the subsequent collapse of the heavy remnant, conserving mass in both evolutionary stages. As discussed above, this updated mass scenario leads to more massive NSs and more LMG BHs, which is illustrated by the broader span of secondary masses and the increased occurrence of binaries with massive NSs and LMG BHs in the right panels of Figure~\ref{fig:binary_mass}. For simulations 1-12, approximately $70\%$ of the primaries in all BH--NS binaries containing an LMG BH ($2.5-5\,\msun$) between 11 and 14~Gyr form from NS--WD mergers, about $25\%$ from NS--NS mergers, and around $5\%$ from NS--MS TDEs. If we also include binaries containing massive NS primaries in the range of $2.1-2.5\,\msun$, then approximately $55\%$ of the massive NSs and LMG BHs form from NS--WD mergers, around $20\%$ from NS--NS mergers, about $10\%$ from NS--MS TDEs, and roughly $10\%$ from WD--WD mergers (not directly forming LMG BHs but massive NSs). Overall, NS--WD mergers contribute the most to the production of massive NS and LMG BH in binaries, followed by NS--NS mergers, in typical core-collapsed GCs similar to NGC~6752. Figure~\ref{fig:form_hist} illustrates the dynamical formation of a NS--NS binary from simulation~12 where the primary NS forms from a NS--WD collision.

\begin{deluxetable*}{ccccc|ccccccc} 
    \tabletypesize{\scriptsize}
    \tablewidth{-1pt}
    \setlength{\tabcolsep}{3pt}
    \tablecaption{Observed and Candidate Pulsar Binaries with NS/BH Companions in Milky Way Globular Clusters} \label{tab:ns_binary}
    \tablehead{
\colhead{Cluster} & \colhead{$M_{clu}$} & \colhead{$r_c$} & \colhead{$r_{hl}$} & \multicolumn{1}{c|}{$N_{\rm PSR}$} & \colhead{PSR} & \colhead{P} & \colhead{$P_{orb}$} & \colhead{$ecc$} & \colhead{$M_{\rm PSR}$} & \colhead{$M_c$} & \colhead{References}\\
\colhead{} & \colhead{[$10^5\msun$]} & \colhead{[arcmin]} & \colhead{[arcmin]} & \multicolumn{1}{c|}{} & \colhead{} & \colhead{[ms]} & \colhead{[day]} & \colhead{} & \colhead{[$\msun$]} & \colhead{[$\msun$]} & \colhead{}
}
    \startdata
    NGC 1851 & 2.8 & 0.09 & 0.51 & 15 & NGC 1851E & 5.596 & 7.448 & 0.708 & $1.53^{+0.18}_{-0.20}$ & $2.35^{+0.20}_{-0.18}$ & (1)\\
    NGC 7078 (M15) & 5.2 & 0.14 & 1.00 & 13 & M15C & 30.529 & 0.335 & 0.681 & $1.358^{+0.01}_{-0.01}$ & $1.354^{+0.01}_{-0.01}$ & (2) \\  
    ${^*}$NGC 1851 & 2.8 & 0.09 & 0.51 & 15 & NGC 1851A & 4.991 & 18.785 &  0.888 & $1.25^{+0.05}_{-0.06}$ & $1.22^{+0.06}_{-0.05}$ & (3) \\
    ${^*}$NGC 6544 & 0.8 & 0.05 & 1.21 & 3 & NGC 6544B & 4.186 & 9.957 & 0.747 &  1.3655 & 1.2064 &  (4)\\
    ${^*}$NGC 7099 (M30) & 1.2 & 0.06 & 1.03 & 2 & M30B & 	12.990 & 6.216 & 0.879 & $\leq 1.43$ & $\geq1.10$ & (5)\\
    ${^*}$NGC 6838 (M71) & 0.38 & 0.63 & 1.67 & 5 & M71D & 100.670 & 10.939 & 0.640 & - & 1.63 & (6)
    \enddata
    \tablecomments{From left to right, the columns show the name of the cluster, cluster mass from https://people.smp.uq.edu.au/HolgerBaumgardt/globular/, the observed cluster core and half-light radii from \citet[][2010 edition]{Harris_1996}, the observed number of pulsars from https://www3.mpifr-bonn.mpg.de/staff/pfreire/GCpsr.html, the name of the pulsar--NS/BH binary or candidate, the pulsar's spin period, binary orbital period, binary eccentricity, mass of the pulsar, mass of the companion, and the reference for the binary properties, respectively. (1) \citet{Barr+2024}; (2) \citet{Jacoby+2006}; (3) \citet{Freire+2004}, \citet{Freire+2007}, and \citet{Ridolfi+2019}  (4) \citet{Lynch+2012}; (5) \citet{Ransom+2004} and \citet{Balakrishnan+2023}; (6) \citet{Pan+2021}, https://www3.mpifr-bonn.mpg.de/staff/pfreire/GCpsr.html, and https://fast.bao.ac.cn/cms/article/65/.\\
    --All of the cluster listed are core-collapsed or nearly core-collapsed (except M71). Note that even though NGC~1851 is shown as non-core-collapsed in \citet[][2010 edition]{Harris_1996}, it has a very small core (e.g., \citealp{Miocchi+2013} and \citealp[][2010 edition]{Harris_1996}) and a very high central density \citep{Baumgardt_Hilker2018}, similar to core-collapsed clusters.\\
    --A $*$ indicates a binary candidate.}
\end{deluxetable*}

Similarly, both NGC~1851-like simulations produce comparable numbers of NS--NS and BH--NS binaries at present (Table~\ref{tab:clu_prop}). Simulation A has zero, while simulation B has about 2 NS/BH--NS per snapshot, roughly consistent with the current pulsar timing observation. Note that NGC~1851E-like pulsar binaries with orbital periods longer than about a week have selection biases similar to other pulsar binaries with massive companions (around $1\,\msun$) and can be detected at around apoastron. Therefore, if a large number of them exist in GCs, they would potentially be detectable.\footnote{However, there are strong selection biases against finding NS--NS binaries with orbital periods shorter than about two days (similar to the M15C binary) due to the effects of large orbital accelerations on pulsar searches. It would be difficult to determine the underlying population of M15C-like binaries in GCs.} 

\begin{figure}
\begin{center}
\includegraphics[width=\columnwidth]{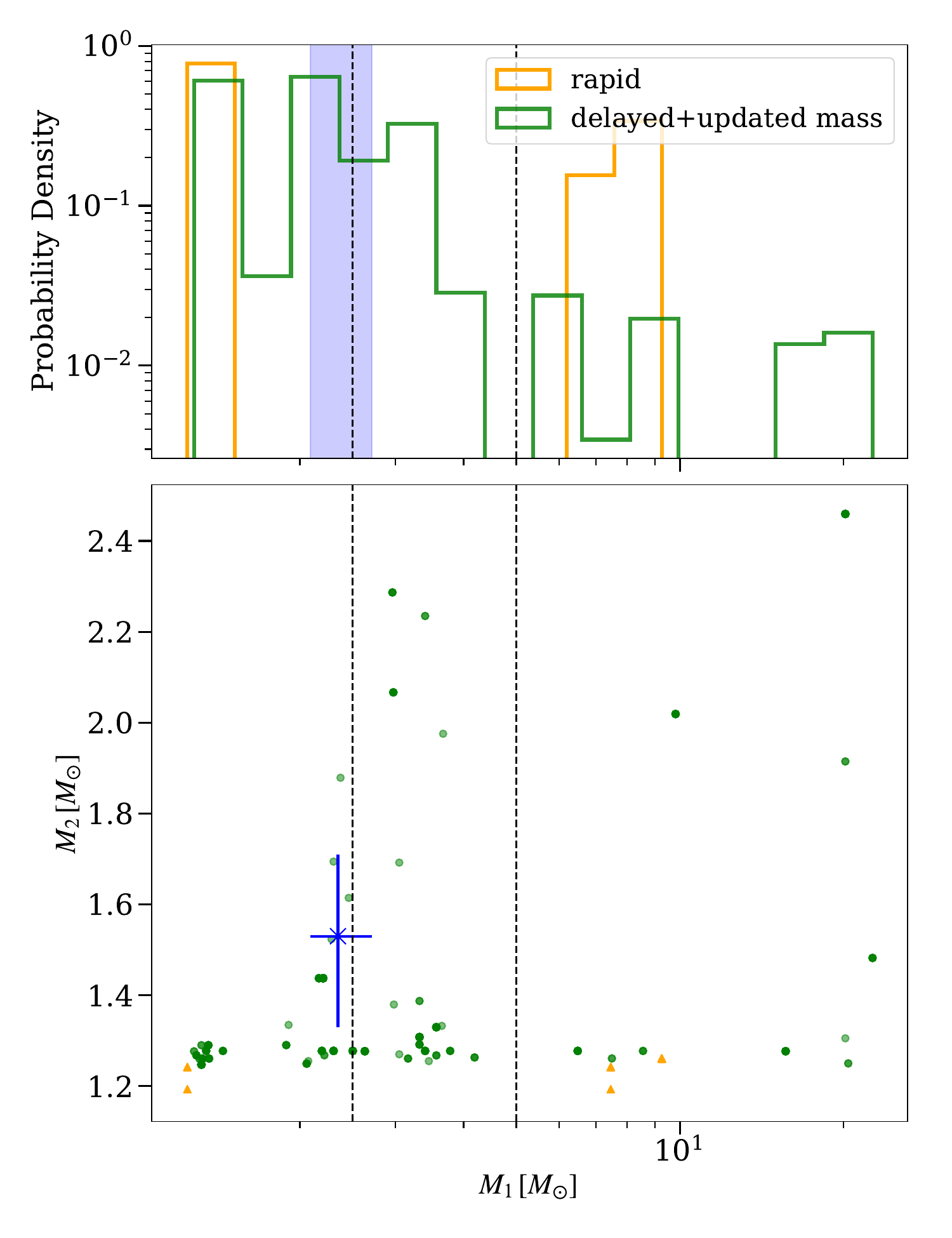}
\caption{Similar to Figure~\ref{fig:binary_mass}, the component masses of all NS--NS and BH--NS binaries (lower panel) and their primary mass distributions (upper panel) between 10 and 13~Gyr in the two NGC 1851-like simulations. The orange triangles and histograms are from simulation A in Table~\ref{tab:clu_prop} with the rapid supernova prescription and the default mass scenario for WD--WD mergers (which does not conserve mass during the mergers). The green circles and histograms are from simulation B with the delayed supernova prescription and the updated WD merger scenario (which conserves mass). The blue cross in the lower panel shows the NGC 1851E binary in Table~\ref{tab:ns_binary}, and the blue band in the upper panel shows the companion mass of NGC 1851E.}\label{fig:1851_binary_mass}
\end{center}
\end{figure}

\begin{figure}
\begin{center}
\includegraphics[width=\columnwidth]{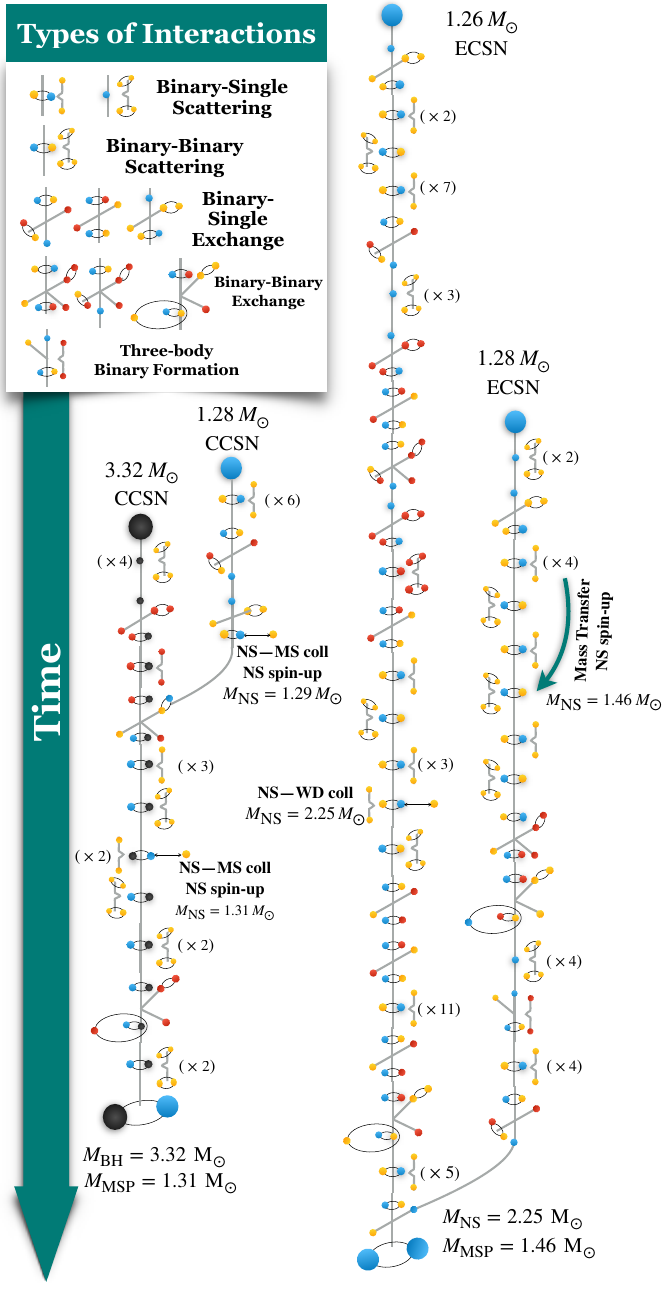}
\caption{Illustration of the dynamical encounters leading to the formation of a LMG BH--NS binary (from simulation B in Table~\ref{tab:clu_prop}) and a NS--NS binary (from simulation 12 in Table~\ref{tab:clu_prop}). Black dots denote the LMG BH and blue dots denote the NSs. On the left, the LMG BH is formed in a delayed supernova (CCSN; a typical formation channel for LMG BHs in massive core-collapsed clusters similar to NGC~1851) and later is paired with a NS through multiple close encounters. The NS in this case is spun up to millisecond periods through NS--MS collisions. On the right, both NSs are formed in electron-capture supernovae (ECSN). One NS collides with a WD and becomes a massive NS with $2.25\,\msun$ (the most likely channel for forming massive NSs and LMG BHs in typical core-collapsed clusters similar to NGC~6752). The other NS accretes about $0.2\,\msun$ through binary mass transfer from one of its binary companions and is spun up to millisecond periods.}\label{fig:form_hist}
\end{center}
\end{figure}

On the other hand, the component mass distribution of the NS--NS and BH--NS binaries differs between the two simulations with different mass scenarios for WD mergers and supernova prescriptions. Figure~\ref{fig:1851_binary_mass} shows the primary and secondary masses of all NS--NS and NS--BH binaries between 10 and 13~Gyr of the clusters' evolution from both NGC 1851-like models. The component masses of the NGC~1851E binary are plotted for comparison. Simulation A produces no BH--NS binaries with primary mass in the LMG. In contrast, for $\sim 50\%$ ($\sim 65\%$) of the time steps between $10$ and $13$~Gyr, the NS--NS or BH--NS binaries in simulation B have primary mass in the range $2.5\,\msun-5\,\msun$ ($2\,\msun-5\,\msun$), including binaries with masses similar to the NGC~1851E binary (bottom panel of Figure~\ref{fig:1851_binary_mass}). Furthermore, simulation B also produces more massive NSs that can better match the mass of NGC~1851E. This figure suggests that the delayed supernova prescription (or, more generally, the formation of LMG BHs from massive star evolution and their retention in dense star clusters) is important for producing LMG binaries in dense and massive clusters similar to NGC~1851. This is because massive clusters can retain more LMG BHs due to their larger escape velocities (Table~\ref{tab:clu_coll_mer}). Furthermore, it also indicates that conserving mass during WD--WD mergers (the updated mass scenario in Section~\ref{sec:cmc}) is a key factor for clusters to produce NGC 1851E-like binaries containing an LMG BH and a massive NS companion \citep[see also, e.g.,][their Figure~9]{Ye+2024_singlemsp}. 

Many NS/BH--NS binaries between 10 and 13~Gyr of simulation B have component masses and orbital properties similar to the NGC~1851E binary. In particular, for binaries with primary masses in the range of $2.09-2.71\,\msun$ and secondary masses in the range of $1.3-1.7\,\msun$ (approximately the same as the components in the NGC~1851E binary), the mean and median eccentricities are about 0.7, and the mean and median orbital periods are about 31 and 13 days, respectively \footnote{Note that these values come from only one simulation. It is possible that the values could change slightly with more simulations using the same initial conditions but different random seeds.}.

Among the NS/BH--NS binaries in simulation B between $10$ and $13$~Gyr whose primary mass is in the range of $2.5\,\msun-5\,\msun$ ($2\,\msun-5\,\msun$), $\sim 30\%$ ($\sim 45\%$) of the time the secondary NS is an MSP (including MSPs formed through NS-main-sequence star TDEs; \citealp{Kremer+2022_nstde,Ye+2024_singlemsp}). We assume that NSs produced in WD--WD mergers are born as young pulsars \citep{Kremer+2023_youngpulsar} rather than MSPs \citep{Ye+2024_singlemsp}. These young pulsars are expected to spin down quickly in $\lesssim 100$~Myr, and are likely undetectable in the present day.

Overall, our NGC~1851-like models can produce NS--NS or BH--NS binaries similar to the NGC~1851E binary. A primary in the LMG most likely forms in a delayed supernova (approximately $75\%$ of all binaries containing an LMG primary between 10 and 13~Gyr) or through the merger between an NS and a massive WD (around $20\%$). Only about $5\%$ of these LMG primaries form from NS--NS mergers. By also including binaries containing massive NS primaries in the range of $2.1-2.5,\msun$, the fraction from the delayed supernovae changes to approximately $55\%$, the fraction from NS--MS TDEs increase to about $5\%$, and WD--WD mergers contribute around $15\%$, with other fractions remaining mostly the same. Thus, the LMG or massive NS companion of NGC~1851E is less likely to be formed in NS--NS mergers. These fractions are very different from those mentioned above for simulations 1-12 due to the difference in the cluster mass and escape velocity. Figure~\ref{fig:form_hist} shows an example of the formation of one of the LMG BH--NS binaries in simulation B.

\section{Gravitational Wave Sources and Luminous Transients with Lower-mass-gap Black Holes}\label{sec:coll_mer}

\subsection{Gravitational Wave Sources}\label{subsec:gwsource}
Some LMG BHs can be paired up with another compact object through dynamical encounters in the clusters and inspiral to produce GW signals in subsequent mergers observable by LVK. 

\begin{figure}
\begin{center}
\includegraphics[width=\columnwidth]{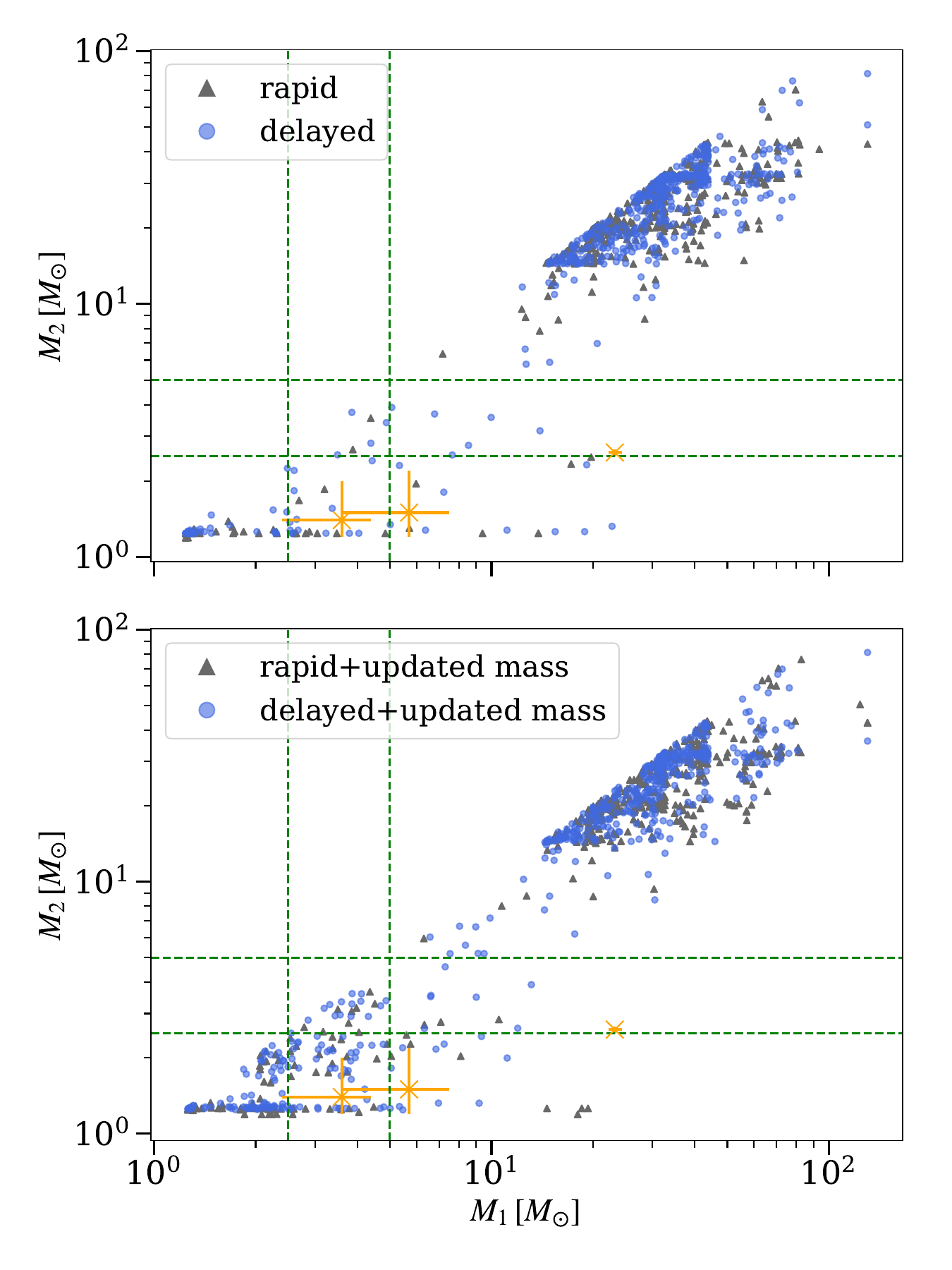}
\caption{Primary and secondary masses for all BH--BH, BH--NS, and NS--NS mergers from simulations 1-12 over a Hubble time. The simulations are divided into four groups similar to Figure~\ref{fig:bh_ns_mass_tot} by their adopted supernova prescriptions and the mass scenarios for WD mergers. The rapid supernova models are shown by black triangles, and blue circles show the delayed supernova models. Simulations in the upper panel assume the default WD mass scenario and the ones in the lower panel assume the updated WD mass scenario. The green lines mark $5\,\msun$ (the assumed upper boundary of the LMG) and the maximum NS mass we adopt at $2.5\,\msun$. GW230529 \citep{GW230529}, GW200115 \citep{GW200115}, and GW190814 \citep{gw190814} are plotted as orange crosses with error bars for comparison. Simulations with the updated mass scenario conserve mass during the mergers of two WDs and the subsequent collapses of the heavy remnants (in contrast to the default scenario). Thus they produce more massive NSs and BHs in the LMG, as well as their mergers.}\label{fig:merger_mass}
\end{center}
\end{figure}

Figure~\ref{fig:merger_mass} shows the component masses of BHs and NSs in all merging binaries over a Hubble time from simulations 1-12. Three GW sources GW230529 \citep{GW230529}, GW200115 \citep{GW200115}, and GW190814 \citep{gw190814} are shown as orange crosses and error bars for comparison. Overall, both supernova prescriptions lead to similar numbers of binary BH mergers ($\sim 130-150$), while the delayed supernova prescription produces slightly more (a factor of $\lesssim 2$) BH--NS and NS--NS mergers (upper panel). This small increase is consistent with most LMG BHs from the delayed supernovae being ejected from the clusters by their natal kicks (Section~\ref{subsec:ccsn}). Simulations with the updated mass scenario for WD mergers form $\sim 2-3$ times more BH--NS and NS--NS mergers compared to those with the default mass scenario (when all other initial conditions and supernova prescriptions are the same) as shown in the lower panel of the same figure. This increase results from the larger number of massive NSs and WDs formed because of the conservation of mass in WD mergers (also see Section~\ref{subsec:nscollapse}). Therefore, simulations with both the delayed supernova prescription and the updated WD mass scenario produce the largest number of NS--NS ($\sim 30$ per core-collapsed cluster up from $\sim 5$) and BH--NS ($\sim 20$ per core-collapsed cluster up from $\sim 5$) mergers. Table~\ref{tab:clu_coll_mer} shows the number of various compact object mergers over a Hubble time and the fraction containing LMG BHs in each simulation.

In general, only a small fraction of binary BH mergers from GCs (a few percent) include LMG BHs between $2.5\,\msun$ and $5\,\msun$. However, $\gtrsim 10\%$ of merging BH--BH from a typical core-collapsed GC can contain at least one LMG BH if NSs are able to accrete very efficiently from TDEs of main-sequence stars and collapse to BHs (simulation 13 in Table~\ref{tab:clu_coll_mer}). These BHs would be fast-spinning and may be detectable by LVK. On the contrary, $\gtrsim 60-70\%$ of BH--NS mergers involve a BH in the LMG for most simulations in Table~\ref{tab:clu_coll_mer}. The recent detection of GW230529, a merger between a LMG BH and an NS from the fourth observing run of LVK, suggests that about $60\%$ of BH--NS mergers may include a component in the LMG \citep{GW230529}. This fraction is similar to what we find for BH--NS mergers in core-collapsed clusters, but the merger rates of GW230529-like systems, $55_{-47}^{+127}\,{\rm Gpc^{-3}\,yr^{-1}}$ \citep{GW230529}, are much larger than the rates expected from GCs. Assuming that about $20\%$ of GCs are core-collapsed (similar to the percentage in the Milky Way) and the GC number density is about $2.3\,{\rm Mpc^{-3}}$ \citep{Rodriguez+2016}, the merger rates of all BH--NS mergers from GCs are roughly $\lesssim 1 \,{\rm Gpc^{-3}\,yr^{-1}}$, even when adopting the delayed supernova prescription and the updated WD mass scenario (the actual rate is probably even lower; \citealp[see][]{Ye+2020_dns}). The extreme upper limit of the BH--NS merger rates is a factor of a few higher and roughly comparable to the BH--BH merger rates, assuming that all NSs and BHs receive small natal kicks, and that not many BHs with masses greater than $10\,\msun$ are formed in clusters (simulation~14). Therefore, even though we can roughly reproduce the observed MSP--LMG BH in NGC~1851, the estimated merger rate of BH--NS is still likely $\gtrsim 2$ orders of magnitude lower than inferred from the most recent LVK data release ($7.8-140\,{\rm Gpc^{-3}\,yr^{-1}}$; \citealp{O3}).

\subsection{Transients from Stellar Collisions and Tidal Disruptions}\label{subsec:transients}
NSs and stellar-mass BHs also dynamically interact with the luminous stars in GCs, as the compact object populations are mixed with the numerous upper main-sequence stars and giant stars in the cluster cores \citep[e.g.,][]{Kremer+2020catalog,Ye+2022_47tuc}. This is directly evidenced by the growing number of observed redback MSPs (https://www3.mpifr-bonn.mpg.de/staff/pfreire/GCpsr.html) and BH binaries with luminous companion stars \citep[e.g.,][]{Giesers+2019} in GCs. Occasionally, the luminous stars pass sufficiently close to the NSs or BHs to be tidally disrupted or directly collide with them. 

We record the number of main-sequence star collisions or TDEs with NSs and BHs over a Hubble time in the simulations in Table~\ref{tab:clu_coll_mer}. In typical core-collapsed clusters with mass of about $2\times10^5\,\msun$ (simulations 1-13), the number of NS--main-sequence TDEs is comparable to the number of BH--main-sequence TDEs, different from the non-core-collapsed clusters where BH TDEs are two orders of magnitude more frequent \citep[e.g.,][]{Kremer+2019_bhtde}. The more massive, NGC~1851-like simulations~A and~B have a smaller number of NS TDEs compared to their BH TDEs, by a factor of about 2--7 depending on the supernova prescription and WD mass scenario. Overall, simulations using the rapid supernova prescription and the default mass scenario for WD merger produce the smallest numbers of TDEs. Among the BH--main-sequence TDEs from simulations 1-12, about $5\%-15\%$ involve a BH in the LMG, with the largest fraction coming from models that adopt the delayed supernova prescription and/or the updated mass scenario. The fractions involving a LMG BH are similar for the NGC~1851-like simulations~A and~B. Furthermore, at most $\sim 20\%$ of BH--main-sequence TDEs can be carried out by LMG BHs in typical core-collapsed clusters if we assume that all mass bound to the NSs during NS--main-sequence TDEs are accreted by the NS, which results in more NSs collapsing to LMG BHs (simulation 13 in Table~\ref{tab:clu_coll_mer}). These TDEs by LMG BHs may give rise to bright X-ray and gamma-ray flares which could resemble ultra-long gamma-ray bursts \citep{Perets+2016}. They may also produce optical signals of luminosity $\sim 10^{41}-10^{44}\,{\rm erg\,s^{-1}}$ that could be detectable by optical transients surveys such as the Zwicky Transient Facility \citep[e.g.,][]{Kremer+2019_bhtde}.

\section{Discussion}\label{sec:discuss}
One of the main uncertainties of the formation and evolution of LMG BHs in dense star clusters lies in the number of systems retained, which is determined by the supernova mechanism and the kicks low-mass BHs and NSs receive at birth. The escape velocity of typical GCs ($\sim 50-100\,{\rm km\,s^{-1}}$) imposes an upper limit on the retention fraction, depending on the natal kick distribution. For NS natal kicks estimated from isolated pulsar observations in the Galactic field \citep[e.g.,][adopted in this study]{Hobbs+2005}, few massive NSs and low-mass BHs (less than about $5\,\msun$) are retained (Table~\ref{tab:clu_coll_mer}), even though $\sim 500$ are formed in a typical GC assuming the delayed supernova mechanism \citep{Fryer+2012}. More recent studies, either based on observed NS binaries \citep[e.g.,][]{Odoherty+2023} or on three-dimensional supernova simulations and semi-analytical models \citep[e.g.,][]{Mandel_Muller_2020}, have instead suggested that NSs may receive smaller natal kicks than previously estimated \citep[e.g., Figure~1 in][and references therein]{Odoherty+2023}. If this is the case, up to $\sim 10-100$ times more NSs may be retained in GCs, considering also that most massive stars are born in binaries \citep{Sana+2012,Moe_DiStefano_2017}. Assuming that LMG BHs, if formed, also receive small natal kicks, an order of magnitude more LMG BHs could be retained in GCs. Together with the larger number of retained NSs, this might lead to a factor of a few more binaries containing LMG BHs in the clusters in the present day. Future (non)observations of LMG binaries will shed more light on the supernova and dynamical formation of these objects in dense star clusters.

We showed that for typical core-collapsed GCs with masses of about $2\times10^5\,\msun$ at present (e.g., NGC~6752), LMG BHs more likely form through dynamical encounters, especially mergers between NSs and WDs. On the other hand, for more massive core-collapsed clusters with present-day masses of about $4\times10^5\,\msun$ (e.g., NGC~1851), core-collapse supernovae through stellar evolution and dynamical interactions (e.g., NS--WD mergers) likely contribute similar numbers of LMG BHs (see also Section~\ref{sec:formation} and Table~\ref{tab:clu_coll_mer}). Additionally, if NSs can accrete nearly all the disrupted stellar debris during TDEs of main-sequence stars, this channel could contribute the most to the formation of LMG BHs. The above suggests that massive, non-core-collapsed clusters may also retain a small number of LMG BHs from massive star evolution.

Another main uncertainty lies in the mass accretion efficiency and mass loss rate of WD--WD mergers. Although zero LMG BHs in the simulations formed directly from the merger-induced collapse of two WDs, it has significant impacts on producing massive NSs, as shown in Section~\ref{sec:binary}. For example, many secondary NSs larger than $1.3\,\msun$ shown in Figure~5 originate from WD--WD mergers, and they are important for matching the mass of NGC~1851E. Significant mass loss during the mergers of two WDs would instead indicate that more massive NSs are formed and retained in the clusters compared to those formed in electron-capture supernovae. This may be due to smaller natal kicks in core-collapse supernovae, as discussed earlier, or because the NSs accrete more mass during previous mass transfer (e.g., through NS--MS TDEs).

The LMG BHs, regardless of their formation channel, could subsequently pair with an MSP or another NS/BH through dynamical encounters, most likely in massive, dense, core-collapsed clusters like M15 and NGC~1851. Core-collapsed clusters probably have only a few BHs remaining \citep[e.g.,][]{Kremer+2018_3201}, allowing NSs to mass segregate to the cluster cores and frequently engage in dynamical interactions \citep[][their Figure~5]{Ye+2019}. Therefore, the formation of LMG BHs through the collapse of NSs and the formation of NS--NS and NS--BH binaries \citep{Ye+2020_dns} are boosted in these dense core-collapsed clusters, and it is not surprising that almost all detected MSP--NS/BH or their candidates are found in (near)core-collapsed clusters (Table~\ref{tab:ns_binary}). 

Similarly, the formation of MSPs \citep{Ye+2019,Ye+2024_singlemsp} and young radio pulsars \citep{Kremer+2023_youngpulsar} is also greatly enhanced in the dense, core-collapsed environment. The large observed number of MSPs (330 pulsars in 44 Milky Way clusters observed to date; https://www3.mpifr-bonn.mpg.de/staff/pfreire/GCpsr.html), although seemingly contradictory, is still consistent with the small number of confirmed NS--NS/LMG BHs in Milky Way GCs (M15C and NGC~1851E). This is because the formation of NS--NS or NS--BH binaries is limited by the dynamical encounter rates of two NSs (single or in binary), which roughly scale with the square of the NS number density in dense star clusters. On the other hand, the formation rate of MSP binaries in dense star clusters scales with the number density of the NSs and the number density of either WDs or luminous stars, which are orders of magnitude higher than the NS number density. For example, the encounter rate between a single and a binary star can be written as
\begin{equation}
    \Gamma_{bs} \approx N_s N_b \Sigma \sigma_v,
\end{equation}
where $N_b$ is the number densities of the target binaries and $N_s$ is the total number of singles of interest. $\Sigma$ is the encounter cross section, and $\sigma_v$ is the velocity dispersion. Thus $\gtrsim 2$ orders of magnitude more observed MSPs in GCs than NS--NS/BH binaries is expected. Likewise, the formation rate of merging NS--NS and BH--NS binaries is also much lower than that of MSPs and the inferred rate from GW detections \citep[see Section~\ref{subsec:gwsource} above; also][]{Ye+2020_dns,O3,GW230529}.

\section{Conclusions}\label{sec:conclu}
In this study, we have explored for the first time the formation and evolution of LMG BHs (in the mass range around $2.5-5\,\msun$) and their binaries through the interplay between massive star evolution and dynamical interactions in dense star clusters. Motivated by the recent detection of a pulsar binary with a potential LMG companion in the massive and dense Milky Way cluster NGC~1851 \citep{Barr+2024}, we model this cluster together with typical core-collapsed GCs using our Monte Carlo code \texttt{CMC}, which self-consistently takes into account stellar (and binary star) evolution as well as all $N$-body dynamics.

Our simulations demonstrate the effects of two main physical mechanisms on the formation of LMG BHs. The first is the supernova core collapse mechanism (i.e., rapid vs delayed collapse; \citealp{Fryer+2012}). The second is dynamical interactions resulting in the collapse of massive compact merger remnants into low-mass BHs (including NS--WD mergers, NS--NS mergers, and tidal disruption of main-sequence stars by NSs). We consider two possible outcomes for mergers of massive WDs (our updated scenario conserves mass, while the default in \texttt{CMC} does not) since WD--WD mergers may produce massive NSs that can later collapse to low-mass BHs in subsequent dynamical interactions.

The delayed supernova prescription produces $\sim 400$ LMG BHs in a typical core-collapsed GC (while the rapid prescription produces none). However, with the large natal kicks they likely receive, only a few LMG BHs are retained in the clusters after formation. More massive GCs similar to NGC~1851 can retain a few tens of these LMG BHs born in delayed supernovae because of the cluster's deeper potential well and the larger number of compact remnants formed initially.

Most of the LMG BHs that form in our models through the collapse of massive NSs originate from NS--WD mergers. In simulations adopting the delayed supernova prescription or the updated WD mass scenario, NS--WD mergers contribute more than twice as many LMG BHs as NS--NS mergers. This is expected since there are many more massive WDs in GCs than NSs and even though most WDs have smaller masses. Only in simulations that adopted the rapid supernova prescription and the default WD mass scenario are there comparable numbers of LMG BHs from NS--WD and NS--NS mergers. In addition, at most a few LMG BHs are produced by NSs tidally disrupting main-sequence stars and accreting the debris. However, if NSs can accrete very efficiently during TDEs (i.e., almost all the mass of the disrupted main-sequence star), then this channel could dominate the production of LMG BHs in core-collapsed clusters \citep[e.g.,][]{Kremer+2022_nstde}.

Among all formation channels, NS--WD mergers can produce $\gtrsim 50\%$ of LMG BHs (with an upper limit of $\sim 85\%$) in typical core-collapsed GCs given the supernova \citep{Fryer+2012} and natal kick prescriptions \citep{Hobbs+2005} adopted in our models. On the other hand, the number of retained LMG BHs from delayed supernovae are comparable to those from NS--WD mergers in more massive core-collapsed clusters similar to NGC~1851. This suggests that LMG BHs formed in supernovae may be significant in more massive clusters, or if they (and NSs) receive smaller natal kicks as indicated by recent studies \citep[e.g.,][]{Mandel_Muller_2020,Odoherty+2023} compared to the previous estimates from isolated pulsar observations \citep[e.g.,][]{Hobbs+2005}. We will explore different supernova and natal kick prescriptions more systematically in future studies.

Our NGC~1851-like models typically have a few BH--NS and NS--NS binaries at present, roughly consistent with the one confirmed system in NGC~1851 (Table~\ref{tab:ns_binary} and \citealp{Barr+2024}) considering that binaries similar to the NGC~1851E binary with orbital periods longer than about a week are not strongly affected by selection biases during pulsar searches. Future (non)detection of BH--NS and NS-NS binaries in GCs would help better constrain supernova models and natal kicks of NSs and LMG BHs. Among the late-time snapshots between 10 and 13~Gyr of the cluster's evolution, about 50\% (65\%) contain NS--NS/BH binaries with primary masses in the range $2.5-5\,\msun$ ($2-5\,\msun$), and about 30\% (45\%) of the time these binaries have an MSP companion. Some of the NS--NS or NS--BH binaries in our models have both component masses similar to those inferred for NGC~1851E. The potential LMG companion of NGC~1851E most likely formed in a core-collapse supernova or a NS--WD merger. Simulations that assume our updated mass scenario, where the total mass is conserved for WD--WD mergers \citep[e.g.,][]{Schwab_2021}, better reproduce the mass of NGC~1851E ($1.53^{+0.18}_{-0.20}\,\msun$).

\begin{acknowledgments}
    We thank the anonymous referee for the helpful comments. C.S.Y. acknowledges support from the Natural Sciences and Engineering Research Council of Canada (NSERC) DIS-2022-568580. Support for KK was provided by NASA through the NASA Hubble Fellowship grant HST-HF2-51510 awarded by the Space Telescope Science Institute, which is operated by the Association of Universities for Research in Astronomy, Inc., for NASA, under contract NAS5-26555. The National Radio Astronomy Observatory is a facility of the National Science Foundation operated under cooperative agreement by Associated Universities, Inc. S.M.R. is a CIFAR Fellow and is supported by the NSF Physics Frontiers Center award 2020265. F.A.R. acknowledges support from NSF grant AST-2108624 and NASA ATP grant 80NSSC22K0722. This research was supported in part through the computational resources and staff contributions provided for the Quest high performance computing facility at Northwestern University, which is jointly supported by the Office of the Provost, the Office for Research, and Northwestern University Information Technology.
\end{acknowledgments}

\vspace{5mm}

\software{\texttt{CMC} \citep{Joshi_2000,Joshi_2001,Fregeau_2003, fregeau2007monte, Chatterjee_2010,Chatterjee_2013b,Umbreit_2012,Morscher+2015,Rodriguez+2016million,CMC1}, \texttt{Fewbody} \citep{fregeau2004stellar}, \texttt{COSMIC} \citep{cosmic}
          }

\bibliography{littleBHs}
\bibliographystyle{aasjournal}

\end{document}